\documentstyle[prb,aps,multicol,epsf]{revtex}
\begin{document}
\draft
\title{Generalized calculation of magnetic coupling constants for
Mott-Hubbard insulators: Application to ferromagnetic Cr compounds}
\author{S. Feldkemper, W. Weber}
\address{Institut f\"ur Physik, Universit\"at Dortmund, 44221 Dortmund,
Germany} 
\date{\today}
\maketitle
\begin{abstract}   
Using a Rayleigh-Schr\"odinger perturbation expansion of multi-band
Hubbard models, we present analytic expressions for the super-exchange
coupling constants between magnetic transition metal ions of arbitrary
separation in Mott-Hubbard insulators. The only restrictions are i) all
ligand ions are closed shell anions and ii) all contributing
interaction paths are of equal length. For short paths, our results
essentially confirm the Goodenough-Kanamori-Anderson rules, yet in
general there does not exist any simple rule to predict the sign of
the magnetic coupling constants.
The most favorable situation for ferromagnetic coupling is found for
ions with less than half filled d shells, the (relative) tendency to
ferromagnetic coupling increases with increasing path length. As an
application, the magnetic interactions of the Cr compounds
Rb$_2$CrCl$_4$, CrCl$_3$, CrBr$_3$ and CrI$_3$ are investigated, all
of which except CrCl$_3$ are ferromagnets.
\end{abstract}
\begin{multicols}{2}
\narrowtext
%
%
%%%%%%%%%%%%%%%%%%%%%%%%%%%%%%%%%%%%%%%%%%%%%%%%%%%%%%%%%%%%%%%%%%%%%%%%%%%%%%%
%
%
\section{Introduction}
The understanding of the magnetic properties of the nonmetallic 3d-transition
metal compounds is not satisfying though they have been investigated
intensively experimentally and theoretically. In particular there
exists only a qualitative understanding of magnitude and sign of the
magnetic coupling on the basis of the so-called Goodenough-Kanamori-Anderson
(GKA) rules. In this paper
we present a more systematic way of calculating the magnetic coupling
constants of these materials based on the concept of
super-exchange. We also test our methods by applying them to a
sub-class of the rare class of ferromagnetic 3d insulators, i.e.~to
Cr compounds, where the GKA rules apparently do not apply. 

In nonmetallic, magnetic 3d-transition metal oxides and halides 
the spins are localized in the incompletely filled 3d shells of the
transition metal sites, hence-forward called magnetic sites.
This is a consequence of large correlation effects. They
are Mott-Hubbard insulators \cite{zaaphd} and are thus described by
a Hubbard-type model of the form
\begin{eqnarray}
\label{hubbard}
H &=&  \sum_{\stackrel{i}{{\alpha},{\sigma}}}
               {\epsilon}^{\alpha}_i \hat{n}_{{\alpha}{\sigma}}(i) +
       \sum_{l,{\sigma}} {\epsilon}_l^{\tau} \hat{n}_{{\tau}{\sigma}}(l)
\nonumber \\ 
  &+&  \sum_{\stackrel{i,l}{{\alpha},{\tau},{\sigma}}}
   [t_{il}^{{\alpha}{\tau}} c^{\dagger}_{{\alpha}{\sigma}}(i)
          c_{{\tau}{\sigma}}(l) + h.c.]   \nonumber  \\
  &+&  \sum_{\stackrel{l,l^{\prime}}{{\tau},{\tau}^{\prime},{\sigma}}}
   [t_{ll^{\prime}}^{{\tau}{\tau}^{\prime}}
   c^{\dagger}_{{\tau}{\sigma}}(l) c_{{\tau}^{\prime}{\sigma}}(l^{\prime}) +
              h.c.]          \nonumber  \\
  &+&  \frac{1}{2} 
       \sum_{\stackrel{\stackrel{i}{{\alpha},{\beta}}}
            {{\sigma},{\sigma}^{\prime}}}   
         U_{{\alpha}{\beta}}\
         c^{\dagger}_{{\alpha}{\sigma}}(i)
         c^{\dagger}_{{\beta}{\sigma}^{\prime}}(i)
         c_{{\beta}{\sigma}^{\prime}}(i)c_{{\alpha}{\sigma}}(i) \nonumber\\
  &-&  \frac{1}{2}       
       \sum_{\stackrel{\stackrel{i}{{\alpha}\neq{\beta}}}
               {{\sigma},{\sigma}^{\prime}}}
         {I}_{{\alpha}{\beta}}\
         c^{\dagger}_{{\alpha}{\sigma}}(i)
         c^{\dagger}_{{\beta}{\sigma}^{\prime}}(i)        
         c_{{\alpha}{\sigma}^{\prime}}(i)c_{{\beta}{\sigma}}(i).
\end{eqnarray} 
Here ${\epsilon}_i^{\alpha}$ is the energy of a d orbital ${\alpha}$ at the
magnetic site $i$ and ${\epsilon}_l^{\tau}$ that of an orbital 
${\tau}$ at the ligand site $l$.
$t_{il}^{{\alpha}{\tau}}$ represents the hopping from the orbital
${\alpha}_i$ to ${\tau}_l$, 
$t_{ll^{\prime}}^{{\tau}{\tau}^{\prime}}$ that from
${\tau}_l$ to ${\tau}^{\prime}_{l^{\prime}}$.
$U_{{\alpha}{\beta}}$ and ${I}_{{\alpha}{\beta}}$ are
Coulomb and  exchange interactions at the magnetic sites, respectively.
In the atomic limit, ${I}_{{\alpha}{\beta}} < 0$ leads to the first
Hund's rule (maximum spin) and $U_{{\alpha}{\alpha}} > U_{{\alpha}{\beta}}$ 
to the second Hund's rule (maximum orbital momentum). 
$\hat{I}_{{\alpha}{\beta}} = {I}_{{\alpha}{\beta}}\,
         c^{\dagger}_{{\alpha}\sigma}(i)
         c^{\dagger}_{{\beta}\sigma^{\prime}}(i)
         c_{{\alpha}\sigma^{\prime}}(i)c_{{\beta}\sigma}(i)
$ will be called  Hund interaction in the following. Its effect is an exchange
of spins between orbitals 
${\alpha}$ and ${\beta}$. The application of $\hat{I}$ to symmetric
(anti-symmetric) spin-states therefore has the effect of a multiplication
with $+(-) I_{{\alpha}{\beta}}$. In Eq.~(\ref{hubbard}) we neg\-lect the
Coulomb interaction at the ligand ions and between different sites. We will
also use the approximation $U_{{\alpha}{\alpha}} = U_{{\alpha}{\beta}} = U$
and ${I}_{{\alpha}{\beta}} =J_H$. These restrictions could be loosened but we
do not expect qualitatively different results. We note that the atomic
part of the Hamiltonian (\ref{hubbard}) has been used very
satisfactorily in ligand field theory for, e.g.~the interpretation of
optical data of such compounds. \cite{sug} 

Anderson has introduced the concept of super-exchange where the hopping of
spins between the different sites leads to the coupling of the total spins
${\bf S}_1$ and ${\bf S}_2$ of the magnetic sites $1$ and $2$. \cite{and1} As
is known from experimental results  \cite{jongh,pick} the interaction is
described very well by the Heisenberg Hamiltonian  
\begin{equation}
H_{\text{eff}} = J_{12}\, {\bf S}_1 \cdot {\bf S}_2.
\end{equation}
The interpretation of
the interaction due to kinetic exchange (different from Coulomb exchange)
is reflected in the formulation of the GKA 
rules. \cite{and} In a simplified version they read
\begin{itemize}
\begin{enumerate}
\item[A)] If the hopping of a spin from a singly occupied ${\zeta}_1$ of the
ion $M_1$ to the singly occupied orbital ${\zeta}_2$ of the ion $M_2$ is
possible, the interaction between the spins in ${\zeta}_1$ and ${\zeta}_2$ is
anti-ferromagnetic.
\item[B)] When a spin can hop from the singly occupied orbital ${\zeta}_1$ to
an unoccupied orbital ${\eta}$ of $M_2$, a ferromagnetic contribution 
to the  interaction results, which is weaker than the anti-ferromagnetic one
from rule A. 
\end{enumerate}
\end{itemize}
With the use of these rules the sign of the coupling constant can be predicted
in many cases. Especially, the GKA rules explain why most of the non-metallic
3d-transition-metal compounds do not show a macroscopic magnetic moment. This
is related to the fact that orbital ordering is usually needed for a 
ferromagnetic interaction to occur. One prominent example are the
CuF$_2$-planes of K$_2$CuF$_4$, where the Jahn-Teller effect causes
orbital order, of alternating $z^2-y^2$ and
$z^2-x^2$ hole orbitals, which completely suppresses the usually dominant
anti-ferromagnetic interaction.

However, in many cases the GKA rules are not helpful to decide which
one of two competing interactions dominates. Such a case is 
Rb$_2$CrCl$_4$, where, again due to the Jahn-Teller effect, an
analogous orbital ordering as in K$_2$CuF$_4$ occurs. Nevertheless,
the nearest neighbor, singly occupied 
$3x^2-r^2$ and $3y^2-r^2$ orbitals exhibit a strong anti-ferromagnetic
coupling via $p_{\sigma}$ ligand orbitals, completely absent
in the case of K$_2$CuF$_4$. 
The fact, therefore, that Rb$_2$CrCl$_4$ is also a ferromagnet, is by
no means a simple consequence of the GKA rules. Similarly, there is no
obvious GKA argument for the fact that the isostructural tri-halides
CrBr$_3$ and CrI$_3$ are ferromagnets.

In our paper we present a more detailed theory of super-exchange
based on the Hamiltonian (\ref{hubbard}) which allows the calculation
of both anti-ferromagnetic and ferromagnetic coupling constants for
hopping paths via arbitrarily many ligands. 

We test our theory by applying it to the ferromagnetic insulators
mentioned above and show that a reasonable choice of parameters in
(\ref{hubbard}) will lead to magnetic coupling constants which agree
with experiment both in magnitude and sign. 

In section II we use perturbation theory to determine the
coupling constants for pairs of d-ions which are linked via an arbitrary number
of ligand ions. First we present the interaction of spin-$\frac{1}{2}$ pairs,
where the spins are exchanged along a chain of ligands each possessing one 
unoccupied orbital. We investigate both the direct super-exchange between
two singly occupied 
d orbitals ${\zeta}$ and the two indirect processes where spin-free (empty or
doubly occupied) d orbitals at the magnetic sites are involved. We then
generalize these results to the case of spin exchange via several paths of
the same order - here we distinguish between equivalent and inequivalent
paths. Finally we give an expression for the coupling of two d-ions with
arbitrary (integral) occupation which are coupled via $P$ different paths
with $N$ unoccupied ligands each. The application of these results to
Rb$_2$CrCl$_4$ and the Cr tri-halides is discussed in section III.

Throughout the paper we will adopt the hole picture since the ligands usually
are anions with a filled p shell. 
%
%
%%%%%%%%%%%%%%%%%%%%%%%%%%%%%%%%%%%%%%%%%%%%%%%%%%%%%%%%%%%%%%%%%%%%%%%%%%%%%%%
%
%
\section{Super-exchange}
In perturbation theory the coupling between the total spins
${\bf S}_1$ and ${\bf S}_2$ of two magnetic ions results from the exchange of
the spins in the singly occupied orbitals. Since we do not consider
spin-dependent interactions like spin-orbit coupling in (\ref{hubbard}), the
spin Hamiltonian is isotropic and in  the lowest order we get
\begin{eqnarray}  
\label{heff1}
H_{\text{eff}}&=& \sum_{{\zeta}_1{\zeta}_2} J_{{\zeta}_1{\zeta}_2}
               {\bf S}_{{\zeta}_1} \cdot {\bf S}_{{\zeta}_2}.
\end{eqnarray}
Here, ${\zeta}_1$ and ${\zeta}_2$ denote the half-filled orbitals at the
sites 1 and 2. Assuming a strong Hund interaction, the ground-state has
maximum spin and the  replacement 
\begin{equation}
{\bf S}_{{\zeta}_i} = \frac{{\bf S}_i}{2\, <{\bf S}_i>}
\end{equation}
appears to be justified. Eq. (\ref{heff1}) can then be written as 
\begin{eqnarray}  
\label{heff2}
H_{\text{eff}} &=& \left (\sum_{{\zeta}_1{\zeta}_2} J_{{\zeta}_1{\zeta}_2} \right ) 
\frac{{\bf S}_1 \cdot {\bf S}_2}{4\,<{\bf S}_1><{\bf S}_2>} \\
&=& J\, \frac{{\bf S}_1 \cdot {\bf S}_2}{4\,<{\bf S}_1><{\bf S}_2>}. 
\nonumber
\end{eqnarray}

We begin with the coupling of spin-$\frac{1}{2}$ pairs. The generalization to
pairs with  larger total spins is possible with  the help of
Eq. (\ref{heff2}).  

We distinguish three cases : \\
$A$: Direct interaction of two singly occupied orbitals ${\zeta}_1$
and ${\zeta}_2$ 
without the involvement of another d orbital. In general, the resulting
coupling will be anti-ferromagnetic, the coupling constant will therefore be
called $J_{{\zeta}_1{\zeta}_2}^A$. \\
$F\!E$: Indirect interaction of ${\zeta}_1$ and ${\zeta}_2$ via one
ore more empty 
d orbitals ${\eta}$ involving the Hund interaction. The coupling constant is
always negative and will be called $J_{{\zeta}_1{\zeta}_2}^{F\!E}$.\\
$F\!D$: Indirect coupling of ${\zeta}_1$ and ${\zeta}_2$ via one ore
more doubly 
occupied d orbitals ${\mu}$ involving the Hund interaction. The coupling
constant is negative, in general, and will be called
$J_{{\zeta}_1{\zeta}_2}^{F\!D}$. 
%
%%%%%%%%%%%%%%%%%%%%%%%%%%%%%%%%%%%%%%%%%%%%%%%%%%%%%%%%%%%%%%%%%%%%%%%%%%%%%%
%
\subsection{super-exchange via one ligand} 
\begin{figure}[p]
\begin{center}
\ \epsfxsize= 8 cm \epsffile{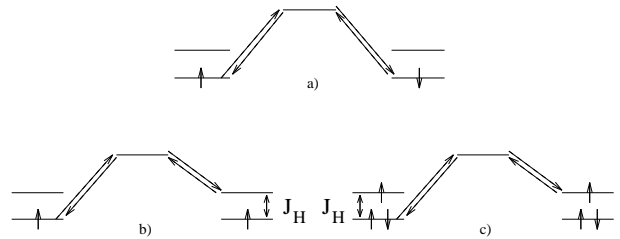}
\end{center}
\narrowtext
\caption{Spin interaction by super-exchange via one empty ligand with one
orbital. a) anti-ferromagnetic case, b) and c) the
ferromagnetic cases. The double arrow represents the direct exchange $J_H$
that leads to the second Hund's rule. b) and c) are differ by the
occupation of the spin-free orbital at the magnetic ion. It is {\em
not} possible to transform one into the other by a particle-hole
transformation.} 
\label{supex}
\end{figure}

We first consider the direct coupling of two equivalent magnetic ions
1,2 with singly occupied orbitals ${\zeta}_1$, ${\zeta}_2$ via one
ligand with an empty hole orbital, see Fig.~\ref{supex} a). Fourth
order perturbation theory is the lowest contributing order. There exist six
possibilities of exchanging the spins which differ in the sequence of spin
exchange. In the following these sequences will be called channels. For
each channel three intermediate states $m$ exist. Two of them contain only
singly occupied sites (one  spin at the ligand and the other at one of the
magnetic ions). In the third state the two spins occupy the same ion. The
excitation energy of the singly occupied states is the charge transfer energy
${\Delta}_{{\zeta}_1}$ (${\Delta}_{{\zeta}_2}$) when the magnetic ion 1 (2) is
singly occupied. We have ${\Delta}_{{\zeta}_i} = {\epsilon}^{\tau} -
{\epsilon}^{\zeta}_i$, when the ligand is empty, which is the usual
case in the hole picture. Note, that when transforming to the electron
picture, the numerical value of ${\Delta}_{{\zeta}_i}$ remains
unchanged, but the orbital energies ${\epsilon}$ have to be
renormalized. If one of the 
magnetic ions is doubly occupied, the excitation energy is  $U \pm
({{\Delta}_{{\zeta}_1}-\Delta_{{\zeta}_2}})$, if it is the ligand we have
${{\Delta}_{{\zeta}_1}+\Delta_{{\zeta}_2}}$. The matrix elements between the
states are the hopping integrals given by (\ref{hubbard}) and each
channel yields 
the same factor $(t_{{\zeta}_1})^2 \, (t_{{\zeta}_2})^2$.  Note that the
energy denominators $({E_0-E_m})^{-1}$ 
are negative, but an additional minus sign occurs because of the {\em
exchange} of the spins.  

The total contribution of the six channels is
\begin{eqnarray} \label{ja11} 
J_{{\zeta }_1{{\zeta }_2}}^{A,1}&=&\left( \frac{1}{{\Delta }_{{\zeta }_1}^2} \,
\frac{1}{U-{\omega }_{{{\zeta }_1}{{\zeta }_2}}} +
\frac{1}{{\Delta}_{{\zeta}_2}^2} \,
\frac{1}{U+{\omega}_{{{\zeta}_1}{{\zeta}_2}}} \right. \\ 
&+&
\left. \frac{1}{{\Delta}_{{\zeta}_1}+{\Delta }_{{\zeta }_2}} \left(
\frac{1}{{\Delta}_{{\zeta }_1}} + \frac{1}{{\Delta }_{{\zeta }_2}} \right)^2
\right) t_{{\zeta }_1}^2 \, t_{{\zeta }_2}^2, \nonumber 
\end{eqnarray}
with 
${\omega }_{{{\zeta }_1}{{\zeta }_2}}={\Delta }_{{\zeta }_2} -
{\Delta}_{{\zeta }_1}$. $t_{{\zeta }_1}$ ($t_{{\zeta }_2}$) is the hopping
integral between orbitals ${{\zeta }_1}$ (${{\zeta }_2}$) and the ligand.

Assuming ${\Delta}_{{\zeta}_1} = {\Delta}_{{\zeta}_2} = {\Delta}_L$
in Eq. (\ref{ja11}), we obtain the well known expression \cite{geer}
\begin{equation}
J^{A,1}_{{\zeta}_1{\zeta}_2} = 2\,
\frac{t_{{{\zeta}_1}}^2\,t_{{{\zeta}_2}}^2}{{{\Delta}_L}^2} \left (
\frac{1}{U} + \frac{1}{{\Delta}_L} \right ). 
\end{equation}
With the replacements
\begin{mathletters}
\label{ja1}
\begin{equation}
2\,\left(\frac{1}{U} + \frac{1}{{\Delta}_L}\right) = A_1 
\end{equation}
\begin{equation}
\frac{t_{{{\zeta}_1}}\,t_{{{\zeta}_2}}}{{{\Delta}_L}} =
\tilde{t}^1_{{\zeta}_1{\zeta}_2}  
\end{equation}
\end{mathletters}
this leads to 
\begin{equation}  \label{JA1}
J^{A,1}_{{\zeta}_1{\zeta}_2} = A_1 \,
\left(\tilde{t}_{{\zeta}_1{\zeta}_2}\right)^2. 
\end{equation}
$\tilde{t}^1$ represents the effective hopping from one to the other
magnetic ion via the ligand. $J^A$ is always non-negative and leads,
because of the Pauli principle, to
an anti-parallel ordering of ${\bf S}_1$ and ${\bf S}_2$.

For the indirect coupling of the spins at the magnetic sites, an
intermediate state with a total spin larger than that of the 
ground state has to be generated at one of the magnetic ions. Thus, there
has to exist at least one spin-free orbital (empty or doubly occupied)
at this ion, which is the
case for most 3d ions except those in high spin $d^5$ configuration. Then an
inter-orbital on-site exchange can take place, leading to a negative
coupling constant.  
 
The intermediate high spin state can occur in two ways. $F\!E$: If the
spin-free orbital at the magnetic ion, where the direct exchange takes
place, is empty, the spin from the half-filled orbital of the other
magnetic ion can hop into it. $F\!D$: If the spin-free
orbital is doubly occupied, one of the spins can hop to
the empty ligand, energetically preferred is the one anti-parallel to
the ground state spin. Both cases lead to a polarization of the ligand
which is parallel to the spin of the magnetic ion in case $F\!E$ and
anti-parallel to it in the case $F\!D$. We will begin with the investigation
of $F\!E$.   

We assume a system of two magnetic ions $M_1$ and $M_2$ with only one singly
occupied orbital ${\zeta}$ and one empty orbital ${\eta}$ each. Also, the
intermediate ligand is empty, see Fig.~\ref{supex} b). 

There exist only two possibilities for exchanging the spins. For each, the spin
from the half filled orbital ${\zeta}$ of one magnetic ion has to migrate to
the empty orbital ${\eta}$ of the other ion, where the on-site exchange of the
spins takes place, leading to the preference of the ferromagnetic
state. Simplifying ${\Delta}_{{\zeta}_1}={\Delta}_{{\zeta}_2}=
{\Delta}_{{\zeta}}$ and ${\Delta}_{{\eta}_1}={\Delta}_{{\eta}_2}=
{\Delta}_{{\eta}}$ the coupling constant is given by
\begin{eqnarray}
J^{F\!E,1}_{{\zeta}_1{\zeta}_2} &=& \frac{J_H}{(U-({\Delta}_{\zeta} -
{\Delta}_{\eta}))^2\, ({\Delta}_{\zeta})^2} \\
&&\left [ (t_{{\zeta}_1} \, t_{{\eta}_2})^2 + 
(t_{{\zeta}_2}\,t_{{\eta}_1})^2 \right ] . \nonumber
\end{eqnarray}
$t_{{\zeta}_i}$ ($t_{{\eta}_i}$) represents the hopping-integral from 
the orbital ${\zeta}_i$ (${\eta}_i$) to the ligand. We note that $J^{F\!E}$
is proportional to $J_H$.

Using the simplification ${\Delta}_{\zeta} = {\Delta}_{\eta} =
{\Delta}_L$ we have
\begin{equation}
J^{F\!E,1}_{{\zeta}_1{\zeta}_2} = \frac{1}{U\,{{\Delta}_L}^2}\,\frac{J_H}{U}
\left [ (t_{{\zeta}_1}\,t_{{\eta}_2})^2 + (t_{{\zeta}_2}\,t_{{\eta}_1})^2
\right ] .
\end{equation}
With the substitutions
\begin{mathletters}
\begin{eqnarray}
 F\!E_1  &=& \frac{J_H}{U^2}
\end{eqnarray}
\begin{eqnarray}
 (\tilde{t}_{{\zeta}_i\,{\eta}_j})^2  &=&
          \frac{t_{{{\zeta}_i}}^2\,t_{{{\eta}_j}}^2}{{{\Delta}_L}^2}
\end{eqnarray}
\end{mathletters}
this expression can be written as
\begin{equation}  \label{JFL1}
J^{F\!E,1}_{{\zeta}_1{\zeta}_2} =
J^{F\!E,1}_{{\zeta}_1{\zeta}_2,{\eta}_2}+
J^{F\!E,1}_{{\zeta}_1{\zeta}_2,{\eta}_1} =
F\!E_1\, [(\tilde{t}_{{\zeta}_1\,{\eta}_2})^2+
          (\tilde{t}_{{\zeta}_2\,{\eta}_1})^2].
\end{equation}
$J^{F\!E}$ contains two different
hopping contributions, each of which is independent of the orbital
${\zeta}$ of that site where the Hund interaction takes place.
$J^{F\!E}$ is non-positive and leads to a parallel orientation of the
spins ${\bf S}_1$ and ${\bf S}_2$. 
This mechanism has been proposed by GKA to explain the occurrence of
ferromagnetic coupling in non-metallic compounds. \cite{and} Yet, since
$|F\!E_1| \ll {A_1} $, it is necessary that
$\tilde{t}^1_{{\zeta}_1{\zeta}_2}$ of Eq.~\ref{ja1} 
disappears or is very small at least. 

For case $F\!D$ we consider a system of two equivalent magnetic ions
$M_1$ and $M_2$ with one singly occupied orbital ${\zeta}$ and one
doubly occupied orbital ${\mu}$ each, see Fig.~\ref{supex} c). As
above, the ligand is empty. 

In contrast to case $F\!E$ there are now 22 ($= 2 \cdot 11$)
channels, instead of two, all leading to a ferromagnetic coupling. The main
difference between 
the channels is whether the spin of ${\zeta}$ or one of the spins of ${\mu}$
first hops to the ligand. Putting
${\Delta}_{{\zeta}_i}= {\Delta}_{{\zeta}}$ and
${\Delta}_{{\mu}_i}= {\Delta}_{{\mu}}$
\begin{eqnarray}  \label{jfd1c}
J^{F\!D,1}_{{\zeta}_1{{\zeta}_2}} &=& J_H\, \left [ \frac{2}
{{\Delta}_{\zeta}\,{\Delta}_{\mu}\,({\Delta}_{\zeta}+{\Delta}_{\mu})} \left (
\frac{1}{{\Delta}_{\mu}}+\frac{1}{{\Delta}_{\zeta}+{\Delta}_{\mu}} \right
)\right. 
\nonumber \\
&+& \frac{1}{({\Delta}_{\mu})^2({\Delta}_{\zeta}+{\Delta}_{\mu})} \left ( 
\frac{2}{{\Delta}_{\mu}}+\frac{1}{{\Delta}_{\zeta}+{\Delta}_{\mu}} \right ) 
 \\
&+& \frac{1}{({\Delta}_{\mu})^2(U+{\Delta}_{\zeta}-{\Delta}_{\mu})} \left ( 
\frac{2}{{\Delta}_{\mu}}+\frac{1}{U+{\Delta}_{\zeta}-{\Delta}_{\mu}} \right )
\nonumber \\
&+& \left. \frac{1}{({\Delta}_{\zeta})^2({\Delta}_{\zeta}+{\Delta}_{\mu})^2}
\right ] \left [ (t_{{\zeta}_1}\,t_{{\mu}_2})^2 +
(t_{{\zeta}_2}\,t_{{\mu}_1})^2 \right ] .  \nonumber 
\end{eqnarray}

A further simplification is obtained when we put ${\Delta}_{\zeta} =
{\Delta}_{\mu} = {\Delta}_L$. Then 
\begin{eqnarray}  \label{JFD1}
J^{F\!D,1}_{{\zeta}_1{\zeta}_2} &=& J_H\,\left[ \frac{1}{{\Delta}_L^3}
\left( \frac{3}{{\Delta}_L}+\frac{2}{U} \right) + \frac{1}{{\Delta}_L^2\,U^2}
\right]\\
&& \left [ (t_{{\zeta}_1}\,t_{{\mu}_2})^2 +
(t_{{\zeta}_2}\,t_{{\mu}_1})^2 \right ] \nonumber \\ 
&=& J^{F\!D,1}_{{\zeta}_1{\zeta}_2,{\mu}_2} +
J^{F\!D,1}_{{\zeta}_1{\zeta}_2,{\mu}_1}  \nonumber \\ 
&=& F\!D_1\, \left [ (\tilde{t}_{{\zeta}_1{\mu}_2})^2 +
(\tilde{t}_{{\zeta}_2{\mu}_1})^2 \right].  \nonumber 
\end{eqnarray}
Here, we have put
\begin{mathletters}
\begin{eqnarray}
F\!D_1 &=&  \frac{J_H}{U} \left(
\frac{1}{U}+\frac{2}{{\Delta}_L}+\frac{3\,U}{{\Delta}_L^2} \right)
\end{eqnarray}
\begin{eqnarray}
(\tilde{t}_{{\zeta}_i{\mu}_j})^2 &=& 
\frac{t_{{{\zeta}_i}}^2\,t_{{{\mu}_j}}^2}{{{\Delta}_L}^2}.  
\end{eqnarray}
\end{mathletters}
$J^{F\!D,1}$ is non-positive and therefore leads to a parallel
orientation of the spins.
The expression is quadratic in the effective hopping-integrals $\tilde{t}$ and
the factor $F\!D_1$ is positive ($U,{\Delta}_L >0$). The coupling constants
$J^{F\!D,1}_{{\zeta}_1{\zeta}_2,{\mu}_i}$ are independent of the orbital
$\zeta$ of the Hund interaction site.
We note that $F\!D_1 > F\!E_1$. As a consequence, the ferromagnetic interaction
should be more favorable for cases of a doubly occupied d orbital than for
those of empty ones. This has already been pointed out
by Anderson. \cite{and2} 
%
%%%%%%%%%%%%%%%%%%%%%%%%%%%%%%%%%%%%%%%%%%%%%%%%%%%%%%%%%%%%%%%%%%%%%%%%%%%%%%
%
\subsection{super-exchange via $N$ ligands}
In the following we consider the interaction via $N$ unoccupied ligands with
one orbital each. We will assume hopping only between nearest
neighbors along the chain of $N+2$ atoms and we investigate again the
cases $A$, $F\!E$ and $F\!D$. 

\begin{figure}[p]
\begin{center}
\ \epsffile{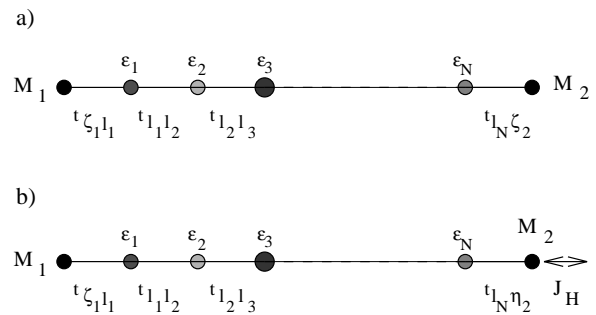}
\end{center}
\narrowtext
\caption{a) Direct interaction on a linear chain between two half-filled
orbitals ${\zeta}_1$ and ${\zeta}_2$ at the sites $M_1$ and $M_2$ via
$N$ unoccupied ligands with one orbital each. Only nearest
neighbor hopping is considered.
b) Indirect interaction between two half-filled orbitals ${\zeta}_1$ and
${\zeta}_2$ at the sites $M_1$ and $M_2$ involving one unoccupied orbital
${\eta}$ via $N$ unoccupied ligands with one
orbital each which build a linear chain. Only the hopping between nearest
neighbors is allowed.}
\label{afnlig}
\end{figure}

Case $A$  requires a
perturbation expansion up to order $2(N +1)$. There are
$\left(^{2\,(N+1)}_{\ N+1}\right)$ possibilities (channels) to
exchange the spins. To take account of all channels we use a simple
method to generate them. For this purpose each channel will be written
as a sequence of numbers 
\begin{equation}
\left( p_1\,p_2\ldots p_{N+1}|p_{N+2} \ldots p_{2N+1}\,p_{2N+2} \right)
\end{equation}
which represents the order of the hopping. Here $p_1$ stands for the hopping
from $M_1$ to the first ligand, $p_2$ for the hopping from the first to the
second ligand, $p_{N+1}$ for the hopping from the last ligand to $M_2$,
$p_{N+2}$ for the hopping back from $M_2$ to the $N$th ligand, see
Fig.~\ref{afnlig}. The arrangement of the numbers in this sequence has to
fulfill the restriction that the left as well as the right half must contain
numbers of increasing magnitude. In the case of an expression of sixth order
the numbers 1 to 6 have to be distributed. Therefore a possible sequence is
\begin{equation}
\left( 1\,2\,4|3\,5\,6 \right).
\end{equation}
First a spin from $M_1$ hops to the first ligand (1) and from there to $l_2$
(2). Then the spin from $M_2$ hops to $l_2$ (3) such that the two spins meet
at $l_2$. Next $M_2$ becomes singly occupied again (4). The last two
processes are the hopping from $l_2$ to $l_1$ (5) and then back to $M_1$
(6). Not possible is the sequence
\begin{equation}
\left( 3\,5\,6|2\,1\,4 \right),
\end{equation}
since the ligands are empty and thus the hopping
from $l_2$ to $l_1$ cannot occur first.

With the simplification ${\Delta}_{{\zeta}_1} = {\Delta}_{{\zeta}_2}$ we find
\begin{equation}
J_{{\zeta}_1{\zeta}_2}^{A,N} = A_N\, (\tilde{t}^N_{{\zeta}_1{\zeta}_2})^2.
\end{equation}
Here 
\begin{mathletters}  \label{an}
\begin{eqnarray}
A_N  &=&2\,\left( \frac{1}{U} + \sum_{k=1}^{N}
\frac{1}{{\Delta}_{l_k}} \right) 
\end{eqnarray}
\begin{eqnarray}
(\tilde{t}^{N}_{{\zeta}_1{\zeta}_2})^2 &=&\frac{
(t_{{\zeta}_1l_1})^2(\prod_{k=1}^{(N-1)} t_{l_kl_{k+1}})^2 
(t_{l_N{\zeta}_2})^2 } {(\prod_{k=1}^{N} {\Delta}_{l_k})^2}.
\end{eqnarray}
\end{mathletters}
The structure of Eq.~\ref{an} is the same as that of Eq.~\ref{JA1} for
ligand. Again, $J^{A,N}$ is the product of a non-negative
hopping-term and a positive energy factor $A_N$. Thus we have
anti-ferromagnetic coupling. Comparing $A_N$ with $A_1$ indicates that
$J^{A,N}$ is reduced as compared to $J^{A,1}$, because
$\tilde{t}^N << \tilde{t}^1$. For large $N$ this is partly compensated
by $A_N > A_1$. 

Case $F\!E$ involving an empty
$M$ orbital, can be handled in the same way. Nevertheless the structure of
the  sequences is somewhat different since the Hund interaction has to be taken
into account. The sequences contain $2\,N + 3$ numbers which are ordered as
follows 
\begin{equation}
\label{fseq}
\left( p_1\,p_2\ldots p_{N+1}|p_{N+2}|p_{N+3}\ldots\,p_{2N+3} \right).
\end{equation}
The central number indicates when the Hund interaction at a specific
magnetic ion (either $M_1$ or  $M_2$), takes place, see Fig.\ \ref{afnlig}.  
There is an additional restriction for
the ordering of the $p_{\lambda}$. Both the series $p_1\ldots p_{N+1}$ and
$p_{N+3}\ldots p_{2N+3}$ as well as the triple $p_{N+1}|p_{N+2}|p_{N+3}$
have to contain numbers of increasing size. For this reason there is only one
channel left. When two ligands are present this is 
\begin{equation}
\left( 1\,2\,3|4|5\,6\,7 \right).
\end{equation}
Then, the interaction constant is given by
\begin{equation}
J_{{\zeta}_1{\zeta}_2}^{F\!E,N} = J_{{\zeta}_1{\zeta}_2,{\eta}_2}^{F\!E,N} +
J_{{\zeta}_1{\zeta}_2,{\eta}_1}^{F\!E,N} = F\!E_N\, \left [
(\tilde{t}^N_{{\zeta}_1{\eta}_2})^2 + (\tilde{t}^N_{{\zeta}_2{\eta}_1})^2
\right]. 
\end{equation}
With ${\Delta}_{\zeta} \approx {\Delta}_{\eta}$ we have
\begin{mathletters}
\begin{eqnarray}
F\!E_N &=& F\!E_1 = \frac{J_H}{U^2} 
\end{eqnarray}
\begin{eqnarray}
\tilde{t}^N_{{\zeta}_i{\eta}_j} &=& \frac{ t_{{\zeta}_il_1}
\prod_{k=1}^{(N-1)} t_{l_kl_{k+1}}\,t_{l_N{\eta}_j} } {\prod_{k=1}^{N}
{\Delta}_{l_k}}.  
\end{eqnarray}
\end{mathletters}
$J_{{\zeta}_1{\zeta}_2,{\eta}_1}^{F\!E,N}$ is non-positive and,
because of $F\!E_N = F\!E_1$, it is much smaller than
$J_{{\zeta}_1{\zeta}_2,{\eta}_1}^{F\!E,1}$ for large $N$, since 
$\tilde{t}^N << \tilde{t}^1$. There is no compensation from an increasing
number of channels. 

Case $F\!D$: Here we find a rapidly growing number of channels. In
sixth order these are 142. For this reason we restrict our
investigation to the case of equivalent (${\Delta}_{l_k} = {\Delta}\
\forall k$) ligands in the chain.  

The sequences introduced in Eq.~\ref{fseq} can also be used here. Yet
the restrictions are modified. The series $p_1\ldots p_{N+1}$ and 
$p_{N+3}\ldots p_{2N+3}$ have to contain numbers of increasing
magnitude, while the center triple $p_{N+1}|p_{N+2}|p_{N+3}$ now contains
numbers of descending size. 

In this situation the coupling constant is (using ${\Delta}_{\zeta}
\approx {\Delta}_{\mu} \approx {\Delta}$)
\begin{eqnarray}
J^{F\!D,N}_{{\zeta}_1{\zeta}_2} &=& J^{F\!D,N}_{{\zeta}_1{\zeta}_2,{\mu}_2} +
J^{F\!D,N}_{{\zeta}_1{\zeta}_2,{\mu}_1}\\
&=& F\!D_N\, \left[
(\tilde{t}^N_{{\zeta}_1{\mu}_2})^2 + (\tilde{t}^N_{{\zeta}_2{\mu}_1})^2
\right]. \nonumber 
\end{eqnarray}
Here
\begin{mathletters} \label{fdn}
\begin{eqnarray} 
F\!D_N &=& \frac{J_H}{U} \left[ \frac{1}{U}+2N\,
\frac{1}{\Delta}+N(2N+1)\,\frac{U}{{\Delta}^2} \right] 
\end{eqnarray}
\begin{eqnarray} 
\tilde{t}^N_{{\zeta}_i{\mu}_j} &=& 
\frac{t_{\zeta_i\,l_1}\,\prod_{k=1}^{(N-1)} t_{l_kl_{k+1}}
t_{l_N\,{\mu}_j}}{{\Delta}^N}. 
\end{eqnarray}
\end{mathletters}

$J^{F\!D,N}$ is non-positive as is $J^{F\!D,1}$. As it contains
terms of order $N^2$ it will compensate the decreasing
magnitude of the hopping terms $\tilde{t}^N$, even more than for the
case $A$. 

The structure of $F\!D_N$ can be made plausible by comparison with $A_N$. The
perturbation series that was needed to calculate $J^A$ has the order $2N +2$,
when $N$ ligands are present, i.e.\ it contains products of $2N+2$
factors. The 
corresponding ferromagnetic interaction additionally includes the
Hund interaction in first order. This factor can be introduced at $2N +1$
positions into the product of ($2N+2$) factors. In case of a $U$-channel in
$A_N$, i.e.\ when the spins meet at a magnetic ion, there is only one
term of order $U^{-2}$, i.e.\ for $p_{N+2} = 2N+1$.. The other $2N$ cases lead
to terms of order $({U\,\Delta})^{-1}$. Then there are $N$ ligand
channels, here each of the $2N+1$ terms
give the same contribution of order ${\Delta}^{-2}$.  

We have restricted the discussion of the coupling via more than one ligand
to the case of ligands of the same type, i.e.~they are either all
anions or all cations. The generalization to the case of mixed-type
paths remains work to be done.
%
%%%%%%%%%%%%%%%%%%%%%%%%%%%%%%%%%%%%%%%%%%%%%%%%%%%%%%%%%%%%%%%%%%%%%%%%%%%%%%
%
\subsection{super-exchange via various paths}
\begin{figure}
\begin{center}
\ \epsffile{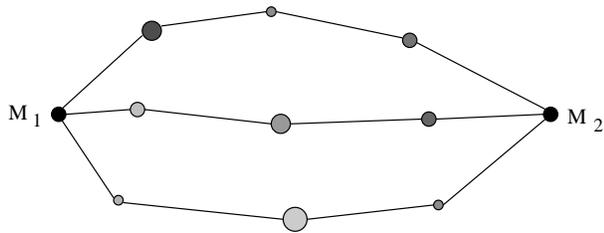}
\end{center}
\narrowtext
\caption{Super-exchange via three different paths with three ligands each.}
\label{ppfad}
\end{figure}

So far we have discussed the super-exchange via one path. But clearly several
paths of the same order can exist along which the spins can be exchanged. A
schematic view of this situation is shown in Fig.\ \ref{ppfad}.

We distinguish between equivalent and non-equivalent paths. Paths $p$ and
$p^{\prime}$ are called equivalent when $Q^p = Q^{p^{\prime}}$ holds, with
$Q^p = A_N^p$, ${F\!E}_N^p$ or $F\!D_N^p$. The additional paths can arise by
introducing further ligand orbitals or by adding further ligand atoms.  

The interaction constant for a multi-path system is calculated by 
combining the contributions of the individual paths. The exchange takes
place in two different ways. The first one is the use of only one path, forth
and back, as before. The second one is a ring exchange using two different
paths forth and back. \cite{carr}

Again we consider the direct coupling of two half filled orbitals for the
cases $A$, $F\!E$ and $F\!D$. Altogether we find the following expressions
\begin{mathletters} \label{npath}
\begin{eqnarray} 
J_{{\zeta}_1{\zeta}_2}^{A,N}(P) &=&\frac{1}{2} \sum_{p,{p^{\prime}} =1}^P
(A^p_N+A^{p^{\prime}}_N)\, \tilde{t}_{{\zeta}_1{\zeta}_2}^{N,p}\,
\tilde{t}_{{\zeta}_1{\zeta}_2}^{N,p^{\prime}} 
\end{eqnarray}
\begin{eqnarray}
J_{{\zeta}_1{\zeta}_2}^{F\!E,N}(P) &=&\frac{1}{2} \sum_{p,{p^{\prime}} =1}^P
(F\!E^p_N+F\!E^{p^{\prime}}_N)\\
&& \left[ \tilde{t}_{{\zeta}_1{\eta}_2}^{N,p}\,
\tilde{t}_{{\zeta}_1{\eta}_2}^{N,p^{\prime}} + 
\tilde{t}_{{\zeta}_2{\eta}_1}^{N,p}\,
\tilde{t}_{{\zeta}_2{\eta}_1}^{N,p^{\prime}} \right]  \nonumber
\end{eqnarray}
\begin{eqnarray}
J_{{\zeta}_1{\zeta}_2}^{F\!D,N}(P) &=&\frac{1}{2} \sum_{p,{p^{\prime}} =1}^P
(F\!D^p_N+F\!D^{p^{\prime}}_N)\\
&& \left[ \tilde{t}_{{\zeta}_1{\mu}_2}^{N,p}\, 
\tilde{t}_{{\zeta}_1{\mu}_2}^{N,p^{\prime}} +
\tilde{t}_{{\zeta}_2{\mu}_1}^{N,p}\,
\tilde{t}_{{\zeta}_2{\mu}_1}^{N,p^{\prime}} \right]. \nonumber
\end{eqnarray}
\end{mathletters}
$P$ denotes the number of paths.

If $Q^p = Q \ \forall p$ the corresponding expressions for $J_Q$ are quadratic
in the effective hopping-integrals.
\begin{mathletters} \label{neqpath}
\begin{eqnarray} 
J_{{\zeta}_1{\zeta}_2}^{A,N}(P) &=& A_N\,\left(\sum_{p=1}^P \,
\tilde{t}_{{\zeta}_1{\zeta}_2}^{N,p} \right)^2 
\end{eqnarray}
\begin{equation}
J_{{\zeta}_1{\zeta}_2}^{F\!E,N}(P) = F\!E_N \left[ \left(\sum_{p=1}^P 
\tilde{t}_{{\zeta}_1{\eta}_2}^{N,p} \right)^2 + \left(\sum_{p=1}^P
\tilde{t}_{{\zeta}_2{\eta}_1}^{N,p} \right)^2 \right] 
\end{equation}
\begin{equation}
J_{{\zeta}_1{\zeta}_2}^{F\!D,N}(P) = F\!D_N \left[ \left(\sum_{p=1}^P 
\tilde{t}_{{\zeta}_1{\mu}_2}^{N,p} \right)^2 + \left(\sum_{p=1}^P
\tilde{t}_{{\zeta}_2{\mu}_1}^{N,p} \right)^2 \right] 
\end{equation}
\end{mathletters}
In all above cases the existence of several equivalent paths only leads to a
new effective hopping $T^{P,N}_{{\alpha}{\beta}} = \sum_{p=1}^P
\tilde{t}^{N,p}_{{\alpha}{\beta}}$. In particular the sign of the coupling
constant remains unchanged, though the effective hopping $T$ and thus the
coupling may vanish when an appropriate set of parameters is chosen.  

In case where the different paths are inequivalent this can lead to a
ferromagnetic coupling resulting from direct super-exchange as has been
discussed in a previous paper. \cite{pap1}  

%
%%%%%%%%%%%%%%%%%%%%%%%%%%%%%%%%%%%%%%%%%%%%%%%%%%%%%%%%%%%%%%%%%%%%%%%%%%%%%%
%
\subsection{total interaction}
The total interaction $J$ of two magnetic ions with total spins ${\bf S}_1$ and
${\bf S}_2$ is obtained from Eq.\ (\ref{heff2}) as the sum of the interaction
of the half-filled orbitals
\begin{equation}
J = \sum_{{\zeta}_1{\zeta}_2} J_{{\zeta}_1{\zeta}_2},
\end{equation}
where
\begin{eqnarray}
J_{{\zeta}_1{\zeta}_2} &=& J_{{\zeta}_1{\zeta}_2}^A+ \sum_{{\eta}_1}
J_{{\zeta}_1{\zeta}_2,{\eta}_1}^{F\!E}+ \sum_{{\eta}_2} 
J_{{\zeta}_1{\zeta}_2,{\eta}_2}^{F\!E}\\ 
&+& \sum_{{\mu}_1}
J_{{\zeta}_1{\zeta}_2,{\mu}_1}^{F\!D}+ 
\sum_{{\mu}_2} J_{{\zeta}_1{\zeta}_2,{\mu}_2}^{F\!D}. \nonumber
\end{eqnarray}
Here $\sum_{{\eta}_i}$ runs over all relevant unoccupied orbitals of $M_i$,
$\sum_{{\mu}_i}$ over the doubly occupied respectively. Then it follows 
\begin{eqnarray}  \label{Jges}
J &=& \sum_{{\zeta}_1{\zeta}_2} \left[ J_{{\zeta}_1{\zeta}_2}^A +
\sum_{{\eta}_1} J_{{\zeta}_1{\zeta}_2,{\eta}_1}^{F\!E}+ \sum_{{\eta}_2} 
J_{{\zeta}_1{\zeta}_2,{\eta}_2}^{F\!E} \right.\\ 
&+&\left. \sum_{{\mu}_1} 
J_{{\zeta}_1{\zeta}_2,{\mu}_1}^{F\!D}+ \sum_{{\mu}_2}
J_{{\zeta}_1{\zeta}_2,{\mu}_2}^{F\!D} \right] \nonumber \\ 
&=& \sum_{{\zeta}_1{\zeta}_2} J_{{\zeta}_1{\zeta}_2}^A +
N_{{\zeta}_1}\,\sum_{{\eta}_1{\zeta}_2}
J_{{\zeta}_1{\zeta}_2,{\eta}_1}^{F\!E} + 
N_{{\zeta}_2}\,\sum_{{\zeta}_1{\eta}_2}
J_{{\zeta}_1{\zeta}_2,{\eta}_2}^{F\!E} \nonumber \\
&+& N_{{\zeta}_1}\,\sum_{{\mu}_1{\zeta}_2}
J_{{\zeta}_1{\zeta}_2,{\mu}_1}^{F\!D} + 
N_{{\zeta}_2}\,\sum_{{\zeta}_1{\mu}_2} J_{{\zeta}_1{\zeta}_2,{\mu}_2}^{F\!D}.
\nonumber
\end{eqnarray}
The factors $N_{{\zeta}_i}$, denoting the number of singly occupied orbitals
at $M_i$, result from the assumption of an orbital-independent on-site
exchange.

We will use equation (\ref{Jges}) to calculate the coupling constants of two
magnetic ions with arbitrary (integral) occupation which are coupled via
unoccupied ligands in the frame of a Hubbard-model. We note that we are
restricted to the case that all contributing paths are of the same order.
%
%
%%%%%%%%%%%%%%%%%%%%%%%%%%%%%%%%%%%%%%%%%%%%%%%%%%%%%%%%%%%%%%%%%%%%%%%%%%%%%%%
%
%
\section{ferromagnetic compounds}
In this section we apply our theory to selected insulating
ferromagnets, i.e.~to  Rb$_2$CrCl$_4$ and the Cr tri-halides CrCl$_3$,
CrBr$_3$ and CrI$_3$. In none of these cases an obvious argument along
the GKA rules can be found for the occurrence of ferromagnetism.  
%
%%%%%%%%%%%%%%%%%%%%%%%%%%%%%%%%%%%%%%%%%%%%%%%%%%%%%%%%%%%%%%%%%%%%%%%%%%%%%%
%
\subsection{Rb$_2$CrCl$_4$}
This compound is a quasi-two-dimensional ferromagnet with a Curie
temperature $T_C = 52.4$ K. \cite{hutch} Values of $J_{\parallel} =
-1.04\ 10^{-2}$ eV and $ \frac{J_{\perp}}{J_{\parallel}} = 3\ 
10^{-4}$ are reported. \cite{hutch} The moments lie in the
CrCl$_2$-planes and are almost parallel. \cite{hutch,jank} 

The crystal structure is analogous to that of K$_2$CuF$_4$. The parent
structure is of K$_2$NiF$_4$-type, a body centered tetragonal lattice
with $a = b = 5.086$ \AA \ and $c =15.72$ \AA. \cite{jank} The Cl$^-$
ligands form an almost perfect octahedron around the Cr$^{2+}$ ions;
yet, because of the $d^4$ configuration, a strong Jahn-Teller
distortion takes place in order to remove the degeneracy of the
$e_g$-orbitals, singly occupied in Cr$^{2+}$.  
The distortion leads to elongation of the ligand cage along the $y$
and $x$ axes, respectively, and correspondingly to contractions in the
$zx$ and $zy$ planes, respectively.\cite{jank}
The same distortions are found in K$_2$CuF$_4$. \cite{kan} Yet there
is one important difference. While for K$_2$CuF$_4$  
($d^9$ configuration of Cu$^{2+}$) with its
occupied d holes of alternatingly $z^2-x^2$ and $z^2-y^2$
type, the GKA rules can be applied in an obvious manner (the
hopping between those orbitals is identically zero by symmetry, and
ferromagnetism results), no such rules apply for the occupied $e_g$-orbitals of
the Cr$^{2+}$ ions which are of $3y^2-r^2$ and $3x^2-r^2$ type,
respectively. This 
orbital order has been established by neutron measurements. \cite{day}
Consequently, ${\sigma}$-type hopping is possible between neighboring
orbitals. In addition, there exists ${\pi}$-type hopping between the
singly occupied $t_{2g}$-orbitals. All these hopping processes favor
an anti-ferromagnetic coupling. Nevertheless, the planar coupling is
ferromagnetic. We note that K$_2$CuF$_4$ and
Rb$_2$CrCl$_4$ also show ferromagnetic coupling along the
c-direction. 

In the hole picture for Cr$^{2+}$, one of the
$e_g$-orbitals is doubly occupied. The other one is singly occupied
(as are the $t_{2g}$-orbitals) and alternates due to the orbital
ordering. When we define the hole state by
\begin{equation}
g({\theta}) = \cos{\theta}|x^2-y^2> + \sin{\theta}|3z^2-r^2>,
\end{equation}
we have values of ${\theta}_{1,2} = \pm  \frac{\pi}{6}$. Then we find
for the anti-ferromagnetic and the ferromagnetic contributions to the
intra-planar coupling
\begin{mathletters}
\begin{eqnarray}
J_{\parallel}^A &=& \sum_{{\zeta}_1{\zeta}_2}
J_{{\zeta}_1{\zeta}_2}^{A} \nonumber   \\ 
&= & \frac{1}{{\Delta}_{Cl}^2} A_1(Cl) \left(2\,(pd{\pi})^4+
\frac{1}{16}(pd{\sigma})^4\right)  
\end{eqnarray}
\begin{eqnarray}
J_{\parallel}^F &=& N_{{\zeta}_1} \sum_{{\zeta}_2{\mu}_1}
J_{{\zeta}_1{\zeta}_2,{\mu}_1} + N_{{\zeta}_2} \sum_{{\zeta}_1{\mu}_2}
J_{{\zeta}_1{\zeta}_2,{\mu}_2}  \nonumber \\
&=& \frac{3}{2} F\!D_1(Cl) \frac{(pd{\sigma})^4}{({\Delta}_{Cl})^2} .
\end{eqnarray}
\end{mathletters}
Here, ${\zeta}$ denotes the singly occupied d-orbitals $xy$, $yz$, $zx$
and $g({\theta}_{1,2})$. For simplicity we have put ${\Delta}_{\zeta}
=  {\Delta}$, i.e.~we have ignored the level splitting of the $t_{2g}$
and $e_g$ orbitals. The total coupling is given by 
\begin{eqnarray}
J_{\parallel} &= & \frac{1}{{\Delta}_{Cl}^2}\left[ 4\,\left(\frac{1}{U}+
\frac{1}{{\Delta}_{Cl}}\right) (pd{\pi})^4 \right. \\
&+& \left. \frac{1}{8} \left(\frac{1}{U}+\frac{1}{{\Delta}_{Cl}}
\right) (pd{\sigma})^4 \right.\nonumber \\
&+& \left.\frac{3}{2} \frac{J_H}{U}\left(\frac{1}{U}+\frac{2}{{\Delta}_{Cl}}+
3\frac{U}{{\Delta}_{Cl}^2}\right)  (pd{\sigma})^4 \right].  \nonumber
\end{eqnarray}
The first two terms represent $J^A$, where the first one, containing
the factor $(pd{\pi})^4$, represents the coupling due to the
$t_{2g}$ orbitals. When we assume the `canonical' ratio
$pd{\pi}/pd{\sigma} \approx -0.4$, the $t_{2g}$ orbitals contribute
less than $1/3$ to $J^A$. Note that any mixed terms
$(pd{\sigma})^2(pd{\pi})^2$ etc.~do not exist, since the $e_g$ and
$t_{2g}$ orbitals are decoupled by symmetry. The third term represents
$J^F$, which contains a factor $J_H/U \approx -0.1$. Yet, on the
other hand, the geometry of the ${\sigma}$-hopping provides a factor
$3/2$ (compared to $1/8$ in term 2). In addition, the large number of
channels in $F\!D_2$ leads to a reduction of the effective energy
denominator by approximately a factor of 2.

As a consequence, a net ferromagnetic coupling is obtained. This
result is rather stable, when energy and hopping parameters are varied
within reasonable limits. We note that this situation may change, when
further octahedral distortions occur which mix the $t_{2g}$ and $e_g$
orbitals. Probably, there exists an easy axis single ion anisotropy,
favoring spin alignments either along the $y$ or the $x$ axis. The
ferromagnetic nearest neighbor coupling will then produce the almost
complete alignment of spins. 

For the related case of KCrF$_3$, with the same structure of Cr planes
as in Rb$_2$CrCl$_4$ suggesting the same orbital structure of the
Cr ions, the puzzle of ferromagnetic
intra-planar coupling has been studied previously by other
researchers. \cite{kuko,ere2} They have assumed direct hopping between
the magnetic sites and have suggested that values of ${\theta} \approx
\frac{\pi}{4}$ should occur, in contrast to the findings in Rb$_2$CrCl$_4$,
which yield ${\theta} = \frac{\pi}{6}$. \cite{day} Note that the neglect of
ligand orbitals strongly reduces the number of ferromagnetic channels -
there is only one left as $F\!D_0 = F\!E_0$ holds. The coupling $J^A$,
which vanishes only for  ${\theta} = \frac{\pi}{3}$, is relatively
large for ${\theta} = \frac{\pi}{6}$ and wins over $F\!D_0$. A
further,  even more questionable consequence is the fact that,
neglecting the ligands, the energy change due to the octahedral
distortion is of the same order as the magnetic interaction. When the
ligands are included, the energy gain due to the Jahn-Teller
distortion dominates over any magnetic energies. An analogous d$^4$
system with planar ferromagnetic coupling is LaMnO$_3$. Again we think
that simplified coupling estimates neglecting the ligands are not
appropriate. \cite{kusa}

An analysis of the inter-planar coupling has not been performed. On the one
hand our formalism is not applicable to the situation with occupied and
unoccupied ligands. On the other hand the number of spins is large (16 in
the hole picture) such that a numerical calculation is very extensive. But we
expect a similar behavior as in K$_2$CuF$_4$, where a reduction of $J^A$
due to destructive interference with a possible sign change and a more or less
unmodified $J^F$ was obtained.\cite{pap1} Furthermore, for Rb$_2$CrCl$_4$ the
spin-free orbital is doubly occupied and therefore the  
case $F\!D$ holds, which always leads to larger ferromagnetic coupling than
the case $F\!E$. 
%
%%%%%%%%%%%%%%%%%%%%%%%%%%%%%%%%%%%%%%%%%%%%%%%%%%%%%%%%%%%%%%%%%%%%%%%%%%%%%%
%
\subsection{CrHa$_3$ (Ha $=$ Cl, Br, I)}
\begin{figure}[h]
\begin{center}
\  \epsffile{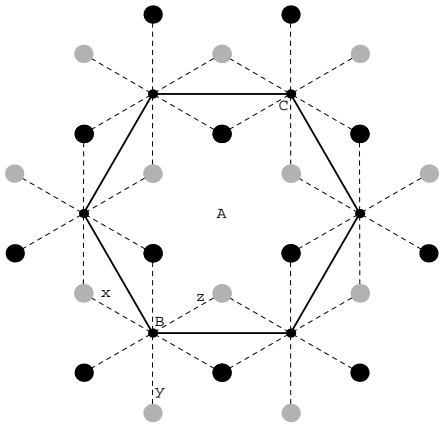}
\end{center}
\narrowtext
\caption{Projection of a Cr-plane with the corresponding ligands on the
$ab$-plane. The small circles denote the Cr-ions, the large ones the
halides. The darker ligand ions lie below the Cr-plane the lighter ones
above. The Cr planes are stacked such that the Cr hexagon center
succesively lies at points A, B and C.}
\label{ebcrha}
\end{figure}

\begin{figure}[tbp]
\begin{center}
\  \epsffile{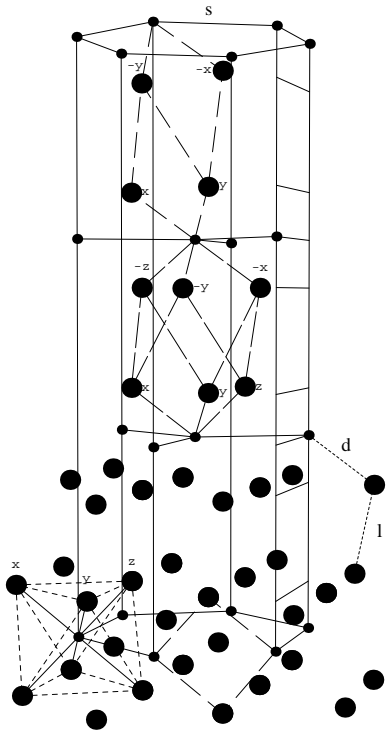}
\end{center}
\narrowtext
\caption{Crystal structure of the tri-halides CrCl$_3$, CrBr$_3$ and
CrI$_3$. The hexagonal unit cell contains four
Cr-planes. Large circles denote the halide ions the smaller circles the Cr
atoms. Clusters corresponding to the
couplings $J_1$, $J_2$ and $J_4$, see
Fig.~\protect\ref{jcrha}, are indicated.  Also shown are the position of a 
ligand octahedron and the corresponding local coordinate system of a
Cr ion. The distances d, l and s are given in 
Tab.~\protect\ref{crdist}.} 
\label{crha}
\end{figure}

The Cr tri-halides with rhombohedral symmetry (R$\bar{3}$) contain two
formula units per trigonal unit cell. The two Cr atoms occupy
equivalent positions, so do all six halide atoms. They form triple
layers Ha-Cr-Ha which are stacked along the c-axis. In the central Cr
layer, the Cr atoms form a honeycomb lattice like a single layer of
graphite. It is helpful to view this plane as a closed packed
triangular arrangement of two Cr atoms (I and II) and of a vacancy V
in the center of the honeycomb. Both the Cr atoms and the vacancy are
surrounded by edge sharing octahedra of halide atoms. As a result, the
two halide planes above and below are closed packed triangular layers
with a stacking sequence  A-B (see Fig.~\ref{ebcrha}). The triple
layers are now repeated along $c$ in such a way that the vacancies V
successively shift by $\frac{1}{3}({\bf a} - {\bf b} + {\bf c})$, where
${\bf a}$ and ${\bf b}$ are the basal plane lattice vectors ($a =
b$, ${\bf a}\cdot {\bf b} = -\frac{1}{2}$) and ${\bf c}$ is the $c$
axis vector of the corresponding hexagonal unit cell. The motion of V
from positions A to B to C to A (see Figs.~\ref{ebcrha} and
\ref{crha}) leads to a packing of the halide planes along $c$ of the
form A-(Cr)-B-(empty)-A-(Cr)-B-(empty)-A-(Cr)-B etc. As a consequence,
the interlayer space B-(empty)-A also consists of halide
octahedra. One can define a perfect octahedral closed packed halide
lattice by requiring that all octahedra are regular (rhombohedral
angle ${\alpha} = 60^{\circ}$). This would lead to the same separation
of A and B planes whether or not a Cr layer is intercalated. It is
clear however that the inter-planar distances are quite different, see
Tab.~\ref{crdist}. The lattice parameters of the Cr tri-halides are given
in Tab.~\ref{stparcrha}, some relevant atomic separations are listed
in Tab.~\ref{crdist}. Note that for our calculations of magnetic
coupling constants it was advantageous to start from an `ideal'
rhombohedral lattice by assuming regular octahedra around the Cr
atoms. In the real lattice, there exists a compression of the octahedra
enclosing the Cr plane. As a result, the octahedra surrounding the
empty layer are considerably elongated along the $c$ direction. There
are two further types of distortion allowed by the R$\bar{3}$
symmetry, which, however, appear to be of minor effect.\newline
1) The Cr planes buckle, i.e.~, the Cr(I) and Cr(II) atoms lie
respectively above and below the plane, as defined by the vacancy
positions (deviation from $1/3$ of the lattice parameter u, see
Tab.~\ref{stparcrha}).
\newline
2) There are some minor shifts in the halide positions (lattice
parameters x deviating from 1/3 or 2/3 values and y deviating from
zero). These shifts result partly from a relaxation of the octahedra
around the vacancy V and partly from the Cr buckling. Note that a
slight rotation of the V octahedra follows, which should lead to
optical activity along the $c$ axis. 

Of the three compounds, the lattice structure of CrBr$_3$ has been
investigated most extensively. \cite{sam,rad} Here, the internal lattice 
parameters x,y,z,u are known, while for CrI$_3$ these
quantities have not been determined. For CrCl$_3$, Morosin and Narath
report a low temperature transition from an orthorhombic to the
R${\bar 3}$ near $T = 225$ K 
and give values for x,y,z,u.\cite{mor} The difference of the two
structures are different stacking arrangements of the Cr tri-halide
planes. To our knowledge, the CrCl$_3$ structural phase transition has
not been confirmed by other groups. There is evidence that
CrCl$_3$ remains orthorhombic to low temperature into the
anti-ferromagnetic phase. \cite{thiele}

All three compounds exhibit strong intra-planar ferromagnetic
coupling. This feature is to be expected from the GKA rules, as the
Cr-Ha-Cr angle for the nearest neighbor (fourth order) super-exchange
paths (leading to the coupling constant $J_1$) is rather close to
$90^{\circ}$ - it is $94^{\circ}$ for CrCl$_3$ and $93^{\circ}$ for
CrBr$_3$.

The inter-planar coupling is anti-ferromagnetic for CrCl$_3$, while it
is ferromagnetic for the other two materials. The critical temperatures
are $T_N^{Cl} = 16.8$ K, $T_C^{Br} = 32.5$ K and $T_C^{I} = 68$
K. \cite{bene} Note that an easy plane anisotropy is reported for
CrCl$_3$, while easy axis anisotropies occur in  CrBr$_3$ and
CrI$_3$. \cite{bene}

From analyses of magnetic susceptibility data intra-planar coupling
constants $J_{\parallel}$ have been obtained. The values reported in
Ref.~\onlinecite{jongh} are $J_{\parallel}^{Cl} = -4.1$ meV,
$J_{\parallel}^{Br} = -6.4$ meV and $J_{\parallel}^{I} =  -10.5$
meV. On the other hand, Samuelsen et al.~have carried out neutron
measurements of spin wave dispersion curves for CrBr$_3$. \cite{sam}
They have been able to extract several near neighbor coupling
constants, in particular for the inter-planar coupling. For the
intra-planar coupling, they report $J_1 = -7.6$ meV and $J_3 = 0.13$
meV (next nearest neighbor, sixth order super-exchange). Note that
$|J_{\parallel}^{Br}| < |J_1 + 2\,J_3|$. Since the $J_1$ value of
Ref.~\onlinecite{sam} probably is the most reliable experimental
quantity available, we have used the relation  $J_1 = 1.2
J_{\parallel}^{Br}$ to scale $J_{\parallel}^{Cl}$ and
$J_{\parallel}^{I}$ accordingly and have assumed $J_1^{Cl} = -4.8$ meV
and $J_1^I = -12.4$ meV. Note that we use the definition of J of
Eq.~\ref{heff2}, leading to a multiplication by a factor of nine of
the values originally given by Refs.~\onlinecite{jongh} and
\onlinecite{sam}.  

\begin{figure}[h]
\begin{center}
\  \epsffile{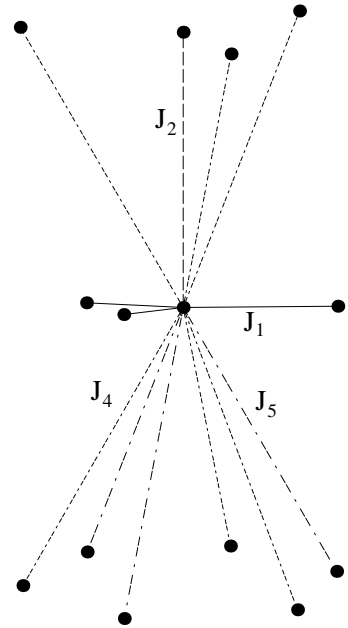}
\end{center}
\narrowtext
\caption{Near neighbor coupling constants $J_1$, $J_2$, $J_4$ and
$J_5$ in CrHa$_3$. There exist
three couplings $J_1$, one $J_2$, three $J_4$  and six $J_5$ per Cr. }
\label{jcrha}
\end{figure}

There exist three different inter-planar coupling constants of sixth
order super-exchange, labeled $J_2$, $J_4$ and $J_5$ by Samuelsen et
al. $J_2$ is the vertical Cr(I)-Cr(II) coupling constant, $J_4$
represents another I-II coupling, while $J_5$ stands for I-I or II-II
couplings (see Fig.~\ref{jcrha}). The values given by Ref.~\onlinecite{sam}
are  $J_2 = 0.22$ meV, $J_4 = -0.06$ meV and $J_5 = -0.12$ meV.

In the following we present results of calculations for the coupling
constants $J_1$, $J_2$, $J_4$ and $J_5$ using the theory of Section
II. We have not included $J_3$, as we mainly focussed on the question
of inter-planar coupling. 

The electronic structure of the Cr ions depends on the ligand
position. Because of almost perfect
octahedra the three singly occupied states are almost $t_{2g}$-like and the
two doubly occupied ones have mostly $e_g$ character, as can be
found from diagonalizing the d-states in the ligand field. Note that
Cr$^{3+}$ is not a Jahn-Teller ion. The three 
empty p-states of the halides are treated as degenerate.  
 
The intra-planar coupling of two nearest Cr ions takes place via two
ligand ions, see Fig.~\ref{crha}. 
From Eq.~(\ref{Jges}) the exchange interaction is given by  
\begin{eqnarray}
J_1 &=& A_1(Ha) \sum_{{\zeta}_1{\zeta}_2}
(T^1_{{\zeta}_1{\zeta}_2})^2 + 6 F\!D_1(Ha) \sum_{{\zeta}{\mu}}
(T^1_{{\zeta}{\mu}})^2  
\end{eqnarray}
with the effective hoppings $T^1_{{\zeta}_1{\zeta}_2}$ and
$T^1_{{\zeta}{\mu}}$ and the energy factors $A_1$ and $F\!D_1$,
Tabs.~\ref{effhopcr}, \ref{hopfaccr}. The 
effective hoppings are calculated corresponding to their definition in
Eq.~(\ref{neqpath}). For the energy denominators,
we have neglected the energy difference between $e_g$ and $t_{2g}$
like states. 

The hopping parameters from Tab.~\ref{crpar} are found by a fitting process, 
where the charge-transfer energy ${\Delta}$ was determined by
adjusting $J_1^{Ha}$ to the experimental values as given
above, while $U$ and $J_H$ were kept fixed. We note that the order of
magnitude and tendency of the ${\Delta }_{Ha}$ and the value for $U$
agree with those given by Zaanen. \cite{zaaphd} These
values are also consistent with those found for Rb$_2$CrCl$_4$.

The calculation of the inter-planar coupling is very complicated. On
the one hand up to six ligands with three orbitals each are involved in
lowest order exchange. On the other hand there are three occupied 
ground state orbitals.

For the calculation of $J_2$, $J_4$ and $J_5$ we need the direct (here
anti-ferromagnetic) and the indirect (ferromagnetic) coupling constant
in sixth order. Each of these couplings is mediated via several
paths. Nevertheless, because we have assumed that all ligand orbitals are
equivalent, this only leads to the introduction of an effective
coupling as was discussed above.  

The vertical inter-planar coupling $J_2$ is mediated via six halide
ions with three orbitals
each. The halide ions form an (empty) octahedron elongated along the $c$ axes,
see Fig.~\ref{crha}. Three halide ions are ligands of the lower Cr ion
and the other three those of the upper one. 
The couplings $J_4$(I-II) and $J_5$(I-I,II-II) are mediated via four
halide ions where to each Cr belong two ligands, see
Fig.~\ref{crha}.

For the coupling constants we get from equations (\ref{neqpath}) and
(\ref{Jges}) 
 \begin{equation}  \label{jicr}
J = A_2(Ha) \sum_{{\zeta}_1{\zeta}_2} (T^2_{{\zeta}_1{\zeta}_2})^2+
6\,F\!D_2(Ha) \sum_{{\zeta}{\mu}} (T^2_{{\zeta}{\mu}})^2.
\end{equation} 
Thus we have to determine the effective hopping integrals
$T^2_{{\zeta}_1{\zeta}_2}$, 
$T^2_{{\zeta}_1{\mu}_2}$ and $T^2_{{\mu}_1{\zeta}_2}$ and the energy factors
$A_2(Ha)$ and $F\!D_2(Ha)$ (independent of the path). 

The terms $T^2_{{\alpha}{\beta}}$ depend on the
individual Cr-halide and halide-halide hopping integrals. In
particular, the ratio $(pp{\sigma})/(pp{\pi})$ of the halide-halide
hopping integrals may modify the relative magnitudes of the $J_2$,
$J_4$ and $J_5$ couplings. 
The values for $T^2_{{\alpha}{\beta}}$ for the inter-planar couplings
are shown in Tab.~\ref{effhopcr} and the resulting energy factors as
well as the values of the $J_2$, $J_4$, $J_5$ coupling constants are
given in Tab.~\ref{hopfaccr}. For CrBr$_3$ these numbers compare well
with the neutron data of Ref.~\onlinecite{sam}, though the ratio
$J_4/J_5$ is not very good. Note that the
`vertical' $J_2$ always is relatively large and anti-ferromagnetic. In
the case of CrBr$_3$ andn CrI$_3$, this coupling is over-compensated by
the ferromagnetic $J_4$ and $J_5$ couplings. This is also found for
CrCl$_3$ unless the ratio $|(pp{\sigma})/(pp{\pi})|$ was shifted from
the usual range of 3 - 5 to an extremely large value of $\approx
36$, such that $J_c = \sum_{i = 2,4,5} z_i\,J_i > 0$. If so, the
inter-planar coupling is probably anti-ferromagnetic for CrCl$_3$ with
this set of parameters. (But note, that $J_c$ can {\em not} be
identified with the inter-planar coupling $J_{\perp}$ in a model with
two couplings $J_{\parallel}$ and $J_{\perp}$. \cite{dav})
Nevertheless, we think that $|(pp{\sigma})/(pp{\pi})| = 36$ is outside
the range of physically 
meaningful parameters. Thus we suspect that there may indeed exist a
difference in the low temperature structures of CrCl$_3$ and the other
two materials. We finally note that the dominant magnetic coupling
constant of these materials is the intra-planar $J_1$ (4th order
super-exchange).  As we pointed out in Sec. II A, the $F\!D$-type
coupling should cause a minority spin polarization on the halide
ligands, as observed in the neutron study of spin density 
distributions in CrBr$_3$ by Radhakrishna and Brown. \cite{rad}
%
%
%%%%%%%%%%%%%%%%%%%%%%%%%%%%%%%%%%%%%%%%%%%%%%%%%%%%%%%%%%%%%%%%%%%%%%%%%%%%%%%
%
%
\section{conclusion}
Nonmetallic transition-metal oxides and halides appear to be the most
prominent members of the family of Mott-Hubbard insulators. In this
paper we have generalized the expansion of extended Hubbard models by
a Rayleigh-Schr\"odinger perturbation series into Heisenberg models,
with the goal to determine general expressions for the super-exchange
coupling constants. Each part
of this series that corresponds to an exchange of the spins in the half-filled
orbitals ${\zeta}_1$ and ${\zeta}_2$ of two magnetic ions $M_1$ and $M_2$
gives a contribution to the (isotropic) coupling constant $J=
\sum_{{\zeta}_1{\zeta}_2} J_{{\zeta}_1{\zeta}_2}$ of the
Heisenberg Hamiltonian 
\begin{equation}  
\eqnum{5}
H = J\,\frac{ {\bf S}_1\cdot{\bf S}_2}
{4\,\langle{\bf S}_1\rangle\langle{\bf S}_2\rangle}.
\end{equation}
Here ${\bf S}_i$ denotes the total spin at the magnetic site $M_i$. This form
of the effective spin Hamiltonian follows from the assumption of a relatively
strong Hund interaction at the magnetic sites such that ${\bf S}_i =
\sum_{{\zeta}_i} {\bf S}_{{\zeta}_i}$.

The couplings $J_{{\zeta}_1{\zeta}_2}$ are built from two different
parts. Either the exchange takes place involving only the d orbitals
${\zeta}_1$ and ${\zeta}_2$. This `direct super-exchange' interaction
is anti-ferromagnetic (A) in general. Or `indirect super-exchange'
occurs, which involves an additional,
empty (E) or doubly occupied (D), d orbital. In this case, because of
the Hund interaction, the coupling is negative in general. These
general results
correspond to the Goodenough-Kanamori-Anderson (GKA) rules as stated
in the introduction, yet quantitative results - and this 
may include the sign of the coupling constant - depend strongly on
details of the involved exchange process. The simplest cases are those
when only one specific path via ligands occurs. Then, each hopping
integral enters quadratically the expressions for the coupling
constants $J$. Therefore, the sign of $J$ is given by kind of
super-exchange, direct ($+$) or indirect ($-$). The magnitude of $J$
depends on i) an effective energy denominator $E_Q$ and ii) on an
effective term $\tilde{t}$, such that $J^{Q} = \tilde{t}^2/E_Q$. Here
$Q= A,F\!E,F\!D$ represents the different possibilities of
super-exchange and $\tilde{t}=(\prod_{k=1}^{N+1} t_k)/{\Delta}^N$,
where $N$ denotes the number of ligands involved in the exchange and
the $t_k$ 
the corresponding hopping integrals. ${\Delta}$ is a characteristic energy of
an excited state with $t_k/{\Delta}\ll 1$. Thus $\tilde{t}$ decreases rapidly
with increasing path length $N$. $1/E_Q$, which we also call $Q_N$,
is strongly influenced by the number of exchange channels. This number
depends on the specific super-exchange $A$, $F\!E$, or $F\!D$, each of
which exhibits a different dependence on $N$. ${F\!E}_N$
is independent of $N$, there are always only two channels. $A_N$ is of the
form  $A_N = 2\,(1/U +N/{\Delta})$ and ${F\!D_N}$ includes in addition a term
quadratic in $N$. For $N=1$ one finds ${F\!E}_1 < {F\!D}_1 < A_1$,
mainly because ${F\!E}_N$ and ${F\!D}_N$ contain an additional factor
$J_H/U$. With growing number of ligands we may find cases $A_N <
{F\!D}_N$. As a consequence, when $N >1$, the interaction of ions with
more than half-filled d-shells 
should exhibit a bigger tendency towards ferromagnetic coupling than
those with less than half-filled d-shells (hole picture, empty
ligands).  

The exchange can also take place via several (P) interfering
paths. The structure 
of a coupling constant resulting from $P$ paths with $N$ ligands
each is  $J^Q(P) = \frac{1}{2} \sum_{p,p^{\prime} = 1}^P(Q_N^p +
Q_N^{p^{\prime}}) \tilde{t}^p\, \tilde{t}^{p^{\prime}}$. Here $p$ and $p'$
indicate the various paths. The energy denominators $Q_N^P$ are the
same as in the case of $P = 1$, yet the hopping factors no
longer enter quadratically into the expressions for the 
effective hopping integrals. In the presence of inequivalent paths ($Q_N^p
\neq Q_N^{p^{\prime}}$) this leads to the possibility of a ferromagnetic
coupling of two spins by direct super-exchange as was discussed in
Ref.~\onlinecite{pap1}. When all $P$ paths are equivalent, the $J^Q$ are
again quadratic in the effective hopping, i.e.~the sign is fixed, but
the magnitude may be strongly influenced by interference. This
behavior always occurs for case $F\!E$, since $F\!E_N$ is independent of $N$.

From the above considerations we see that even the sign of the
coupling cannot be guessed except in the simplest cases, even if the orbital
order at the magnetic sites is known. For better quantitative
estimates, a reasonable knowledge of hopping integrals, of charge
transfer gaps and of $U$ parameters is required. These numbers are
accessible, in principle, via total energy calculations using
constrained density functional theory, see
e.g.~Ref.~\onlinecite{hyb}. Our present studies are based on such
evaluations for La$_2$CuO$_4$. 

In order to test the quality of our treatment, we have investigated
some ferromagnetic Cr compound insulators, where the GKA rules cannot
be applied in any simple fashion. In Rb$_2$CrCl$_4$, with
the same structure as K$_2$CuF$_4$, orbital order of the Cr$^{2+}$
ions occurs because of the Jahn-Teller effect. However, for the first
nearest neighbor (1NN) intra-planar coupling, the direct super-exchange
$J_A$ does not vanish, in contrast to the case of
K$_2$CuF$_4$. Nevertheless, we find $|J_{F\!D}^1| > J_A^1$, for
various reasons as discussed above and in detail in chapter 3. Also
the quantitative result agrees well with experiment. Concerning the
inter-planar coupling, we believe that Rb$_2$CrCl$_4$ is rather
similar to K$_2$CuF$_4$ which has been studied
previously. \cite{pap1} We note that some of the $N = 2$ super-exchange
paths involve both empty (Cl$^-$) and doubly occupied (Rb$^+$)
ligands. Such cases, which may occur in ternary compounds, are not
covered by our present theoretical treatment.
 
The Cr$^{3+}$ ions of the Cr tri-halides CrCl$_3$, CrBr$_3$ and
CrI$_3$ are not Jahn-Teller ions and, thus, do not exhibit orbital
ordering. The 1NN intra-planar 
coupling involves two equivalent $N=1$ paths with Cr-Ha-Cr angles
close to $90^{\circ}$. The $F\!D$ coupling dominates the $A$ coupling,
both due to the $90^{\circ}$ angles and, also because of the very many
$F\!D$ channels. The inter-planar coupling takes place via three
different $N=2$ contributions, $J_2$, $J_4$ and $J_5$. Each of the
couplings is mediated via a variety of paths. Using reasonable hopping
parameters we get good agreement with the values $J_2$, $J_4$, $J_5$
as measured by neutron scattering in the case of CrBr$_3$. Further we
obtain ferromagnetic coupling for both CrBr$_3$ and CrI$_3$. In the case
of CrCl$_3$ we suspect that the low temperature structure is not
R$\bar{3}$. 

We come to the conclusion that the perturbational calculation of the coupling
constants in the non-metallic 3d-transition-metal compounds is an appropriate
instrument, qualitatively as well as quantitatively. The quantitative
agreement may appear somewhat surprising. It has been pointed out earlier
that the higher order contributions of the perturbation series can have a
similar magnitude as the lowest one. \cite{eskes} Under such circumstances we
would expect that the coupling between next nearest neighbors becomes
important. Experimentally, however, ratios $J_2/J_1 \approx 10^{-2}$
are reported. \cite{pick} An additional consequence should be that
terms of the form  
$({\bf S}_i \cdot {\bf S}_j)({\bf S}_k \cdot {\bf S}_l)$ can not be
neglected. We are not aware of any experimental evidence for the
importance of such 4-spin terms. The question
remains how the good agreement between experimental value and the lowest
order results can be understood. 
%
%%%%%%%%%%%%%%%%%%%%%%%%%%%%%%%%%%%%%%%%%%%%%%%%%%%%%%%%%%%%%%%%%%%%%%%%%%%%%%%
%
%                             FIGURES
%
%%%%%%%%%%%%%%%%%%%%%%%%%%%%%%%%%%%%%%%%%%%%%%%%%%%%%%%%%%%%%%%%%%%%%%%%%%%%%%%
%
%
%
%%%%%%%%%%%%%%%%%%%%%%%%%%%%%%%%%%%%%%%%%%%%%%%%%%%%%%%%%%%%%%%%%%%%%%%%%%%%%%%
%
%                             TABLES
%
%%%%%%%%%%%%%%%%%%%%%%%%%%%%%%%%%%%%%%%%%%%%%%%%%%%%%%%%%%%%%%%%%%%%%%%%%%%%%%%
%
%
\begin{table}[tbp]
\narrowtext
\caption{Electronic parameters for Rb$_2$CrCl$_4$ (in eV). }
\label{rbcrclpar}\vspace{2mm} \centering
\ 
\begin{array}[b]{ccccc}
\hline\hline
(pd{\sigma})&(pd{\pi})&U&J_H&{\Delta}_{Cl} \\  
-1.0 &0.4&4&-0.52 &4.15\\   
\hline\hline
\end{array}
\ 
\end{table}
\begin{table}[tbp]
\narrowtext
\caption{Lattice parameters for CrCl$_3$
(Ref.~\protect\onlinecite{rad}),  for CrBr$_3$ (Ref.~\protect \onlinecite{sam})
and  for CrI$_3$ (Ref.~\protect\onlinecite{mor}). For CrI$_3$
the $x$, $y$, $z$ and $u$  values are approximated by the `ideal' values,
$2/3$, $0$, \mbox{$ z_i = a_0/ 2 \, \protect\sqrt{3} c_0 $ }, and
$1/3$, respectively.
} 
\label{stparcrha}\vspace{2mm} \centering
\
\begin{array}[b]{c*{3}{r@{.}l}}
\hline\hline
   &  \multicolumn{2}{c}{\mbox{CrCl}_3}   &
   \multicolumn{2}{c}{\mbox{CrBr}_3}    &
   \multicolumn{2}{c}{\mbox{CrI}_3}\\ 
a_0&  5&94 \mbox{\AA} &  6&27 \mbox{\AA}  &  6&86 \mbox{\AA} \\    
c_0& 17&33 \mbox{\AA} & 18&21 \mbox{\AA}  & 19&88 \mbox{\AA} \\
x  &  0&6507          &  0&6523           &  0&6666           \\
y  & -0&0075          &  0&0012           &  0&0            \\          
z  &  0&0757          &  0&0786           &  0&0813          \\
u  &  0&3323          &  0&3339           &  0&3333           \\
\hline\hline
\end{array}
\
\end{table}
\begin{table}[tbp]
\narrowtext
\caption{Relevant atomic separations for the CrHa$_3$ compounds (in
{\AA}, for definition, see Fig.~\protect\ref{crha}). Because of the
distortions $u \neq 1/3$, $y \neq 0$ etc. there exist two different
distances $d$ (Cr-Ha) and $l$ (Ha-Ha). Also given are the separations
of the Ha planes $r_1$ (Cr enclosed) and $r_2$ (empty site enclosed)
 }
\label{crdist}\vspace{2mm} \centering
\
\begin{array}[b]{cccc}
\hline\hline
&\mbox{CrCl}_3&\mbox{CrBr}_3&\mbox{CrI}_3\\ 
s  &3.43 &3.62 &3.96 \\
d_1&2.37 &2.50 &2.80 \\
d_2&2.33 &2.50 &2.80 \\
l_1&3.69 &3.80 &4.09 \\
l_2&3.80 &3.87 &4.09 \\
r_1&2.62 &2.86 &     \\
r_2&3.15 &3.21 &     \\
\hline\hline
\end{array}
\
\end{table}
\begin{table}[tbp]
\narrowtext
\caption{Electronic parameters (in eV) for CrHa$_3$. The
hopping integrals are given for the `ideal' 
structure. Distance variations are included via exponentials
$e^{|{\beta}|(r_0-r)}$ with ${\beta}^{pd}_{\sigma} = 1.5 $ eV/\AA, 
${\beta}^{pd}_{\pi} = 2 $ eV/\AA , ${\beta}^{pp}_{\sigma} = 2$ eV/\AA \ and
${\beta}^{pp}_{\pi} = 2.5$ eV/\AA, where $r_0$ denotes the distance in
the ideal structure.  }
\label{crpar}\vspace{2mm} \centering
\ 
\begin{array}[b]{c*{3}{r@{.}l}}
\hline\hline 
   &  \multicolumn{2}{c}{\mbox{CrCl}_3}   &
   \multicolumn{2}{c}{\mbox{CrBr}_3}    &
   \multicolumn{2}{c}{\mbox{CrI}_3}\\ 
{\Delta}     & 3&8   & 3&3  & 3&0\\
U            & 3&5   & 3&5  & 3&5\\
\frac{J_H}{U}&-0&1   &-0&1  &-0&1\\
(pd{\sigma}) &-1&15  &-1&12 &-1&06\\
(pd{\pi})    & 0&66  & 0&69 & 0&70\\
(pp{\sigma}) & 0&05  & 0&10 & 0&12\\
(pp{\pi})    &-0&0014&-0&019&-0&04\\
\hline\hline
\end{array}
\ 
\end{table}
\begin{table}[tbp]
\narrowtext
\caption{ 
CrHa$_3$: Values of various effective hopping squares for the 
couplings $J_1$, $J_2$, $J_4$ and $J_5$ (in
$10^{-4}$ eV$^2$). } 
\label{effhopcr}\vspace{2mm} \centering
\ 
\begin{array}[b]{cccc}
\hline\hline
&\mbox{CrCl}_3&\mbox{CrBr}_3&\mbox{CrI}_3\\[2mm]
&\multicolumn{3}{c}{J_1}\\[1mm]
\sum_{{\zeta}_1{\zeta}_2} (T^2_{{\zeta}_1{\zeta}_2})^2 
&9733  &7670 &4802 \\[1mm]  
\sum_{{\zeta}{\mu}} (T^2_{{\zeta}{\mu}})^2 
&12211  &8862&5506\\[2mm]
&\multicolumn{3}{c}{J_2}\\[1mm]
\sum_{{\zeta}_1{\zeta}_2} (T^2_{{\zeta}_1{\zeta}_2})^2 
&84  &359 &444 \\[1mm]  
\sum_{{\zeta}{\mu}} (T^2_{{\zeta}{\mu}})^2 
&12  &98 &182\\[2mm] 
& \multicolumn{3}{c}{J_4} \\[1mm]
\sum_{{\zeta}_1{\zeta}_2} (T^2_{{\zeta}_1{\zeta}_2})^2 
&18 & 79  &107 \\[1mm]  
\sum_{{\zeta}{\mu}} (T^2_{{\zeta}{\mu}})^2 
&18  &88  &140\\[2mm] 
& \multicolumn{3}{c}{J_5} \\[1mm]
\sum_{{\zeta}_1{\zeta}_2} (T^2_{{\zeta}_1{\zeta}_2})^2 
&11 &69  &107 \\[1mm]  
\sum_{{\zeta}{\mu}} (T^2_{{\zeta}{\mu}})^2 
&13  &85  &140 \\
\hline\hline
\end{array}
\
\end{table}
\begin{table}[tbp]
\narrowtext
\caption{CrHa$_3$: Effective energy denominators $A$ and $F\!D$ (in eV$^{-1}$) and coupling constants (in meV).}  
\label{hopfaccr}\vspace{2mm} \centering
\ 
\begin{array}[b]{c*{3}{r@{.}l}}
\hline\hline 
   &  \multicolumn{2}{c}{\mbox{CrCl}_3}   &
   \multicolumn{2}{c}{\mbox{CrBr}_3}    &
   \multicolumn{2}{c}{\mbox{CrI}_3}\\
A_1      & 1&10  & 1&18  & 1&24\\
6\,F\!D_1&-0&94  &-1&11  &-1&28\\
A_2      & 1&64  & 1&77  & 1&91\\
6\,F\!D_2&-2&28  &-2&76  &-3&36\\
J_1      &-4&8   &-7&6   &-12&4\\
J_2      & 0&059 & 0&26  & 0&31\\
J_4      &-0&007 &-0&08  &-0&33\\
J_5      &-0&006 &-0&09  &-0&33\\
\hline\hline
\end{array}
\ 
\end{table}
%
%
%
%%%%%%%%%%%%%%%%%%%%%%%%%%%%%%%%%%%%%%%%%%%%%%%%%%%%%%%%%%%%%%%%%%%%%%%%%%%%%%%
%
%				REFERENCES	
%
%%%%%%%%%%%%%%%%%%%%%%%%%%%%%%%%%%%%%%%%%%%%%%%%%%%%%%%%%%%%%%%%%%%%%%%%%%%%%%%
%

\end{multicols}
\end{document}